\documentclass[aps,twocolumn,pra,superscriptaddress]{revtex4-2}
\usepackage{amsfonts}
\usepackage[final,hiresbb]{graphicx}
\usepackage{amsmath}
\usepackage{amssymb}
\usepackage{amstext}
\usepackage{color}
\usepackage{braket}

\usepackage{subfigure}
\usepackage{appendix}
\usepackage{bbm}

\definecolor{Green}{RGB}{0,204,102}
\definecolor{Purple}{RGB}{102,0,255}
\definecolor{Blue}{RGB}{51,153,255}
\definecolor{Red}{RGB}{255,010,010}

\begin{document}

		\title{Quantized Optical Vortex-Array Eigenstates in a Rotating Frame}

	\author{Mark T. Lusk}
	\email{mlusk@mines.edu}
	\affiliation{Department of Physics, Colorado School of Mines, Golden, CO 80401, USA}
 
	\author{Andrew A. Voitiv}
	\affiliation{Department of Physics and Astronomy, University of Denver, 2112 E. Wesley Avenue, Denver, CO 80208, USA}
	
	\author{Mark E. Siemens}
	\email{Mark.Siemens@du.edu}
	\affiliation{Department of Physics and Astronomy, University of Denver, 2112 E. Wesley Avenue, Denver, CO 80208, USA}

	\begin{abstract}
Linear combinations of Bessel beams can be used to effectively trap light within cylindrical domains. Such hard traps can be used to produce states that exhibit stationary arrays of optical vortices from the perspective of a steadily rotating frame. These patterned singularities can be engineered to have singularities of the same or mixed charges, and the requisite rotation rates are quantized even though the setting is purely linear. A hydrodynamic interpretation is that the vortices are at rest within a compressible, two-dimensional fluid of light.
	\end{abstract}
	\maketitle

	\section{Introduction}

Rotating frames are ubiquitous in physics because they allow for the investigation of phenomena that cannot normally be brought into equilibrium\cite{Landau1996}.  For instance, one of the remarkable features of trapped, rotating Bose-Einstein condensates (BEC) is their ability to support arrays of vortex eigenstates\cite{Yarmchuk1979, Fetter2009}. While many comparisons have been drawn between 
vortices in BEC and singularities generated in coherent beams of light\cite{Coullet1989, Allen1992}, there is no optical analog to such trapped eigenstate arrays. This is addressed in the present work by structuring beams so that energy and information is confined to a fixed radial domain, effectively creating an all-optical trap. These structures are engineered so that the state within such traps appear fixed when viewed from a frame that spirals with a specified pitch along the propagation axis. With this axis interpreted as time, the result is a means of producing optical eigenstates with patterned vortices.

For BEC, a circular hard trap is realized by the steric containment afforded by a vessel or external ``bucket" potential \cite{Gaunt2013, Barenghi2016}.
We show an optical setting that offers a particularly attractive way of accomplishing the same thing. Zeroes of idealized Bessel beams enforce such hard-trap boundary constraints; there is no energy flux across these boundaries, so restricting attention to dynamics within the first Bessel zero is equivalent to evolution within a hard-trap domain. This allows trapped vortex dynamics to be studied in free-space optics. This setting is advantageous, compared to trapped BEC or fiber-trapped light, because the pattern and charge of the optical vortices can be engineered by tailoring the Bessel mode weights using digital holography. Likewise, a thorough characterization of the resulting beam can be carried out using standard optical equipment because of the free-space setting.

When viewed in a frame rotating with appropriate angular speed, these vortex arrays exhibit motion that is dramatically different from dynamics in a free-space Gaussian beam; the presence of the trap admits vortex states that are perfectly stationary and do not appear to interact (although still consistent with hydrodynamics \cite{Andersen2021}). While the behavior is purely classical and linear, the requisite rotation rates are quantized. 

In this paper, a theoretical framework for the construction of such rotating-frame eigenstates is developed and then applied to produce and analyze a variety of charge-neutral eigenstates. These states are then experimentally implemented and characterized, and optical vortex eigenstates are observed in a rotating frame.

	\section{Single-Vortex Eigenstates}

Under a paraxial approximation for the monochromatic electromagnetic vector potential, ${\bf A}(r, \phi, z , t) = {\bf e}_0 A_0\psi(r, \phi ,z)e^{i(k z - \omega t)}$, electrodynamics are governed by a two-dimensional Schr\"odinger equation\cite{Lax1975}:
\begin{equation}\label{paraxial}
	i \partial_z \psi = -\frac{1}{2k} \nabla^2_\perp \psi .
\end{equation}
The associated dynamics are then characterized by a scalar field, $\psi$, with the z-axis treated as time. This field can be confined by restricting attention to a circular domain for which radius $r \le r_0$ and assuming that the field is separable---i.e.
\begin{equation}\label{psiform}
	\psi(r, \phi, z) = u(r) e^{\imath m \phi} e^{-\imath \varepsilon z}.
\end{equation}
The scalar field, $u$, then satisfies Bessel's eigenvalue problem,
\begin{equation}\label{Bessel}
	\partial_{r,r} u + \frac{1}{r} \partial_r u - \frac{m^2}{r^2} u = -2 k \varepsilon u,
\end{equation}
with modes described by Bessel functions of the first kind, $J_m$:
\begin{equation}\label{EVP}
	u(r) = J_m (\sqrt{2 k \varepsilon}r) .
\end{equation}
A hard-trap boundary condition of $u(r_0)=0$ quantizes the admissible eigenvalues to
\begin{equation}\label{eps}
	\varepsilon_{mj} = \frac{\nu_{mj}^2}{2 k r_0}.
\end{equation}
Here $\nu_{m,j}$ is the $j^{th}$ Bessel zero of the Bessel function of order $m$. This delivers a set of mutually orthogonal modes that satisfy the boundary condition at $r=r_0$,
\begin{equation}\label{nu}
	\psi_{mj} (r, \phi) =  J_m (\nu_{mj} r/r_0)  e^{\imath m \phi} ,
\end{equation}
where $m\in \mathbb{Z}$ and $j\in\mathbb{N}$. Each mode is equivalent to a sum of plane waves propagating on the surface of a cone that subtends an angle, $\alpha_{mj}$, with respect to the z-axis\cite{McGloin2005}:
\begin{equation}\label{alpha}
	\alpha_{mj} =  \sqrt{2 \varepsilon_{mj}/k}.
\end{equation}
The enforcement of $\alpha_{mj} \ll 1$ ensures that the paraxial approximation is satisfied\cite{Potocek2015}. Each solution, Eq. \ref{nu}, has a vortex of charge $m$ at the center of the trap. Positive charges correspond to a phase that increases from $0$ to $2\pi$ with counter-clockwise motion around the vortex.

Linear combinations of these Bessel modes can be used to construct evolving states, within the domain $r \le r_0$, for which the field is zero at $r=r_0$:
\begin{equation}\label{init}
	\psi (r, \phi, z) = \sum_{m, j} c_{mj} \psi_{mj} (r, \phi) e^{-\imath \varepsilon_{mj} z} .
\end{equation}
Here constants $c_{mj}$ can be used to satisfy an arbitrary initial condition.

	\section{Eigenstates Composed of Vortex Arrays}

There exists a particularly simple class of dynamics for which a linear combination of stationary Bessel states can itself be a stationary state within a steadily rotating frame. To see this, consider the scalar field of Eq. \ref{init} as viewed in a frame that rotates with a steady angular speed of $\Omega$ about the z-axis:
\begin{equation}\label{rot1}
	\acute\psi(r, \phi, z) := \psi(r, \phi - \Omega z, z) .
\end{equation}
Define an azimuthal variable for this new frame, $\acute \phi = \phi - \Omega z$. Since
\begin{equation}\label{rot2}
	\psi(r, \phi, z) = \acute\psi(r, \acute \phi, z) ,
\end{equation}
we have that 
\begin{equation}\label{rot3}
	\partial_z \psi(r, \phi, z) = \partial_z \acute\psi(r, \acute \phi, z) + \Omega \, \partial_{\acute\phi} \acute\psi(r, \acute\phi, z) .
\end{equation}
Noting that the z-component of the angular momentum operator is $L_z = -\imath \partial_{\acute \phi}$, the paraxial equation in the rotating frame is\cite{Fetter2009}
\begin{equation}\label{rot4}
		i \partial_z \acute\psi = -\frac{1}{2k} \nabla^2_\perp \acute\psi + \Omega L_z \acute \psi.
\end{equation}
Look for separable solutions by assuming that
\begin{equation}\label{sep}
		\acute\psi(r, \acute \phi, z) = u(r, \acute\phi) e^{-\imath \eta z} .
\end{equation}
This gives the following eigenvalue problem:
\begin{equation}\label{evp}
		\biggr[-\frac{1}{2k} \nabla^2_\perp  + \Omega L_z \biggl] u = \eta u .
\end{equation}

Consider a potential solutions to Eq. \ref{evp} of the form
\begin{align}\label{usol}
		u^{(m)}_{j_1,j_2} =& e^{-\imath \alpha/2} \cos(\beta/2) e^{-\imath m \phi} J_m(\nu_{m,j_1}r/r_0) \nonumber \\
		&+e^{\imath \alpha/2} \sin(\beta/2) e^{\imath m \phi} J_m(\nu_{m,j_2}r/r_0 ) ,
\end{align}
where parameters $\alpha$ and $\beta$ control the phase and magnitude of the mode weighting. This will satisfy Eq. \ref{evp} provided that an associated eigenvalue, $\eta^{(m)}_{j_1,j_2}$, and rotation rate, $\Omega^{(m)}_{j_1,j_2}$, are specified as follows:
\begin{equation}\label{eta}
		\eta^{(m)}_{j_1,j_2} = \frac{1}{4 k r_0^2} (\nu_{m,j_1}^2 + \nu_{m,j_2}^2),
\end{equation}
\begin{equation}\label{Omega}
		\Omega^{(m)}_{j_1,j_2} =\frac{\nu_{m,j_1}^2 - \nu_{m,j_2}^2}{4 m  k r_0^2} .
\end{equation}

\begin{figure}[t]
	\begin{center}
		\includegraphics[width=\linewidth]{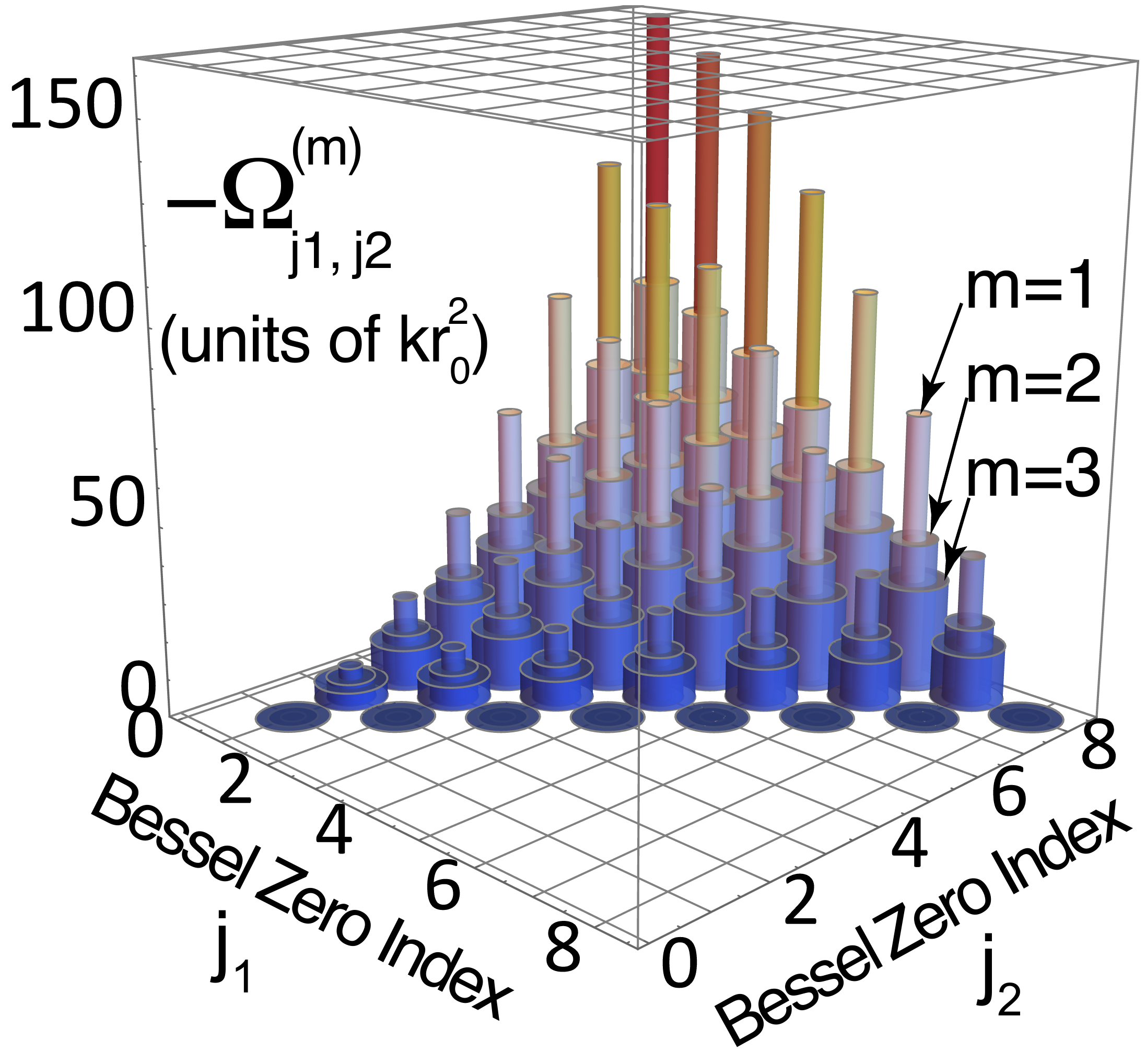}
	\end{center}
	\caption{ \emph{Quantized Rotation Rates in a Bessel Trap}. The quantized rotation rates, $\Omega^{(m)}_{j_1,j_2}$, of Eq. \ref{eta}$_2$, for Bessel zero indices, $j_1$, ranging from 1 to $j_2$, and $j_2$ ranging from 1 to 8. A complementary set of positive rotation rates is associated with states for which $j_1 \ge j_2$. The first three Bessel orders are plotted, $m = 1, 2, 3$. The first three Bessel orders, $m = 1, 2, 3$, are plotted as concentric cylindrical shells. For a given pair of Bessel zero indices, the rotation rate decreases with Bessel order, $m$. } 
	\label{Fig1}
\end{figure}

We now have a class of stationary trapped states within a rotating frame. An immediate observation is that the requisite rotation rates are quantized per Eq. \ref{Omega}. These are plotted in Fig. \ref{Fig1} for a range of Bessel zeroes, $\nu_{m,j_1}$ and $\nu_{m,j_2}$, and the lowest three Bessel orders. As shown there, rotation rate magnitude increases with the difference between the Bessel zero indices, $j_1$ and $j_2$. 

%
%

Each of these quantized rotation rates can be used to produce eigenstates in the rotating frame. In fact, a continuous spectrum of weighting coefficients, $\alpha$ and $\beta$, are associated with each quantized choice of rotation rates, $\Omega^{(m)}_{j_1,j_2}$ and eigenvalues, $\eta^{(m)}_{j_1,j_2}$. 

\section{Polar Arrays of Vortex Eigenstates}

Eigenstates in the rotating frame are expressed as a linear combination of just two Bessel modes of order $m$, but the states exhibit multiple pairs of linear-core, tilted vortices of charge $\pm 1$ along with a linear-core, tilted vortex of charge $m$ at the center of the domain, $r<r_0$. Each eigenstate, Eq. \ref{usol}, is parametrized by Bessel order, $m$, Bessel zero indices $j_1$ and $j_2$, and tilt angles, $\alpha$ and $\beta$. These 5 parameters offer a great deal of control over the structural features of vortex arrays. 

\begin{figure}[t]
	\begin{center}
		\includegraphics[width=0.9\linewidth]{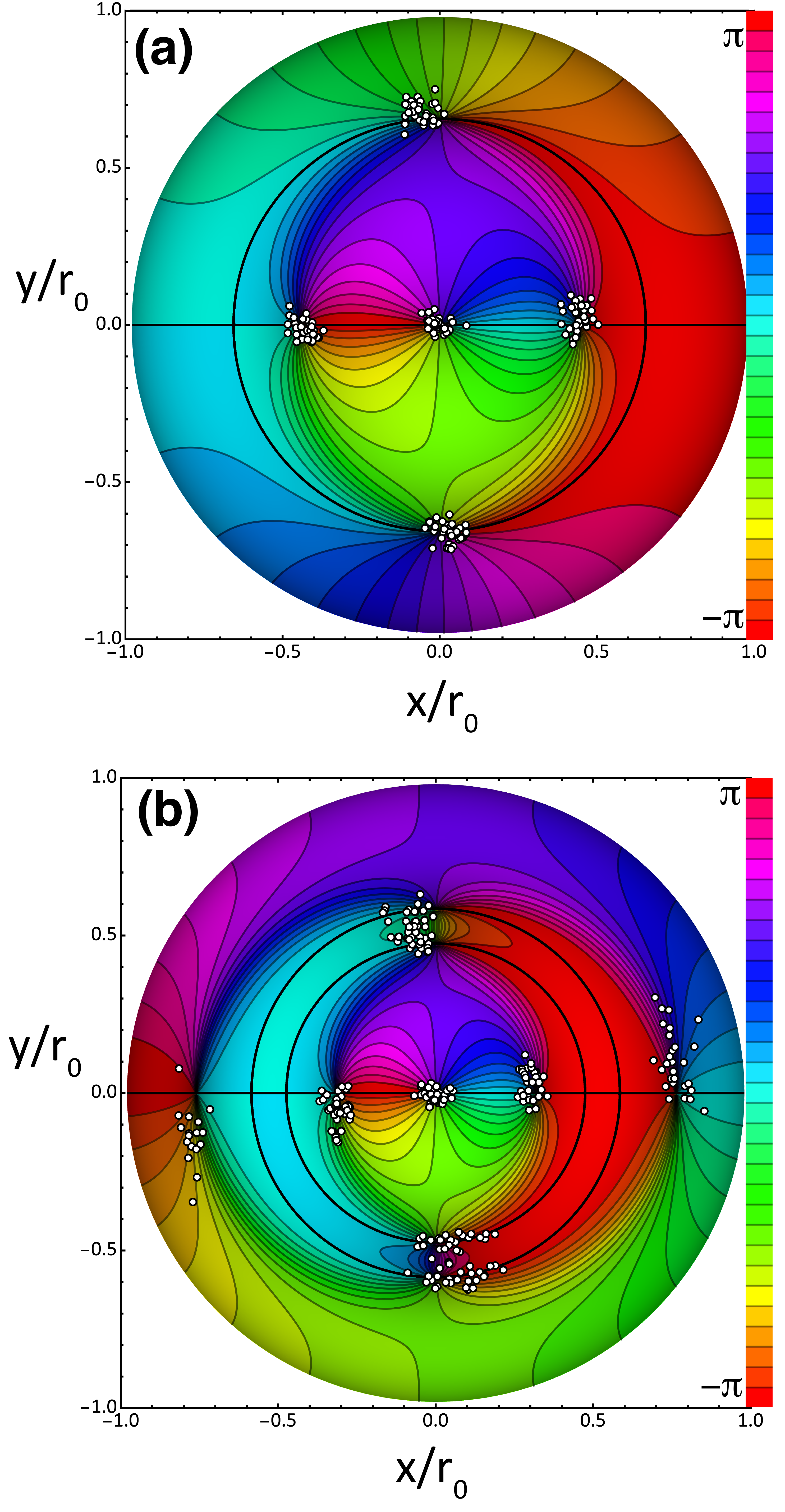}
	\end{center}
	\caption{ \emph{Vortex-Array Eigenstates in a Bessel Trap}.  Frames rotating in accordance with Eq. \ref{eta}$_2$ support a wide range of vortex-array eigenstates, Eq. \ref{usol}, parametrized by Bessel order, $m$, Bessel zero indices $j_1$ and $j_2$, and tilt angles, $\alpha$ and $\beta$. Contour phase plots are shown for two eigenstates for which $m=1$. In both cases, $\alpha=0$ and $\beta=125^\circ$. White dots correspond to experimentally measured vortex positions, as discussed in the text, and correspond to a hard trap radius of $r_0 = 0.4$ mm.  (a) $m=1$, $j_1 = 1$, $j_2 = 2$,  (b) $m=1$, $j_1 = 1$, $j_2 = 3$. } 
	\label{Dipoles_1}
\end{figure}

Two particularly simple arrays are shown in Fig. \ref{Dipoles_1}, for which the Bessel modes are of the lowest order. Panel (a) shows an eigenstate phase and associated vortex array 
consisting of 2 positive and 2 negative vortices arranged about a +1 vortex at the center. The state shown in panel (b) differs from that of (a) by only 1 parameter, but it features a pair of vortex dipoles. In fact, the spacing between the dipoles can be smoothly increased/decreased by increasing/decreasing tilt angle, $\beta$. The existence of eigenstates supporting vortex dipoles is in stark contrast to freely propagating\cite{Andersen2021} and harmonically trapped\cite{Zhu2022} settings where vortices of opposite charge contribute to each others' background field, leading to annihilation\cite{Andersen2021}.

%
%

\begin{figure}[t]
	\begin{center}
		\includegraphics[width=0.9
	\linewidth]{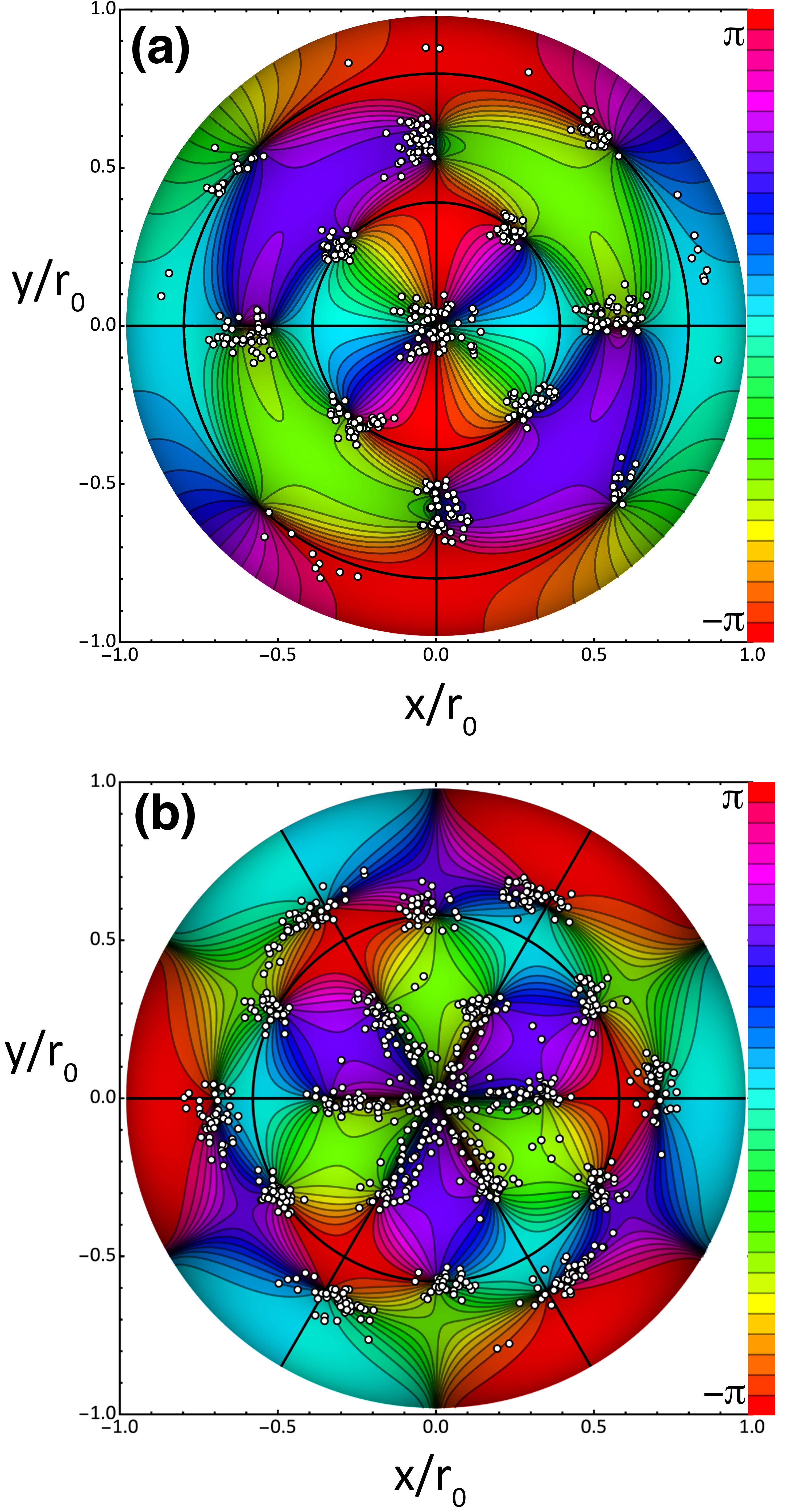}
	\end{center}
	\caption{ \emph{Higher-Order Vortex Arrays}.  Two eigenstates of Eq. \ref{usol} are shown. Contour phase plots are shown for two eigenstates for which $\alpha=0$. White dots correspond to experimentally measured vortex positions, as discussed in the text, and correspond to a hard trap radius of $r_0 = 0.4$ mm. (a) A second-order version of Fig. \ref{Dipoles_1}(b) with $\beta=120^\circ$, $m=2$, $j_1=1$, $j_2=3$.  (b) An array without dipoles for which the inner 12 vortices are arranged on the vertices of equilateral triangles with $\beta=78.565^\circ$, $m=3$, $j_1 = 2$, $j_2 = 3$. 
    }
	\label{Result_Composite_2}
\end{figure}

\begin{figure}[t]
	\begin{center}
		\includegraphics[width=1\linewidth]{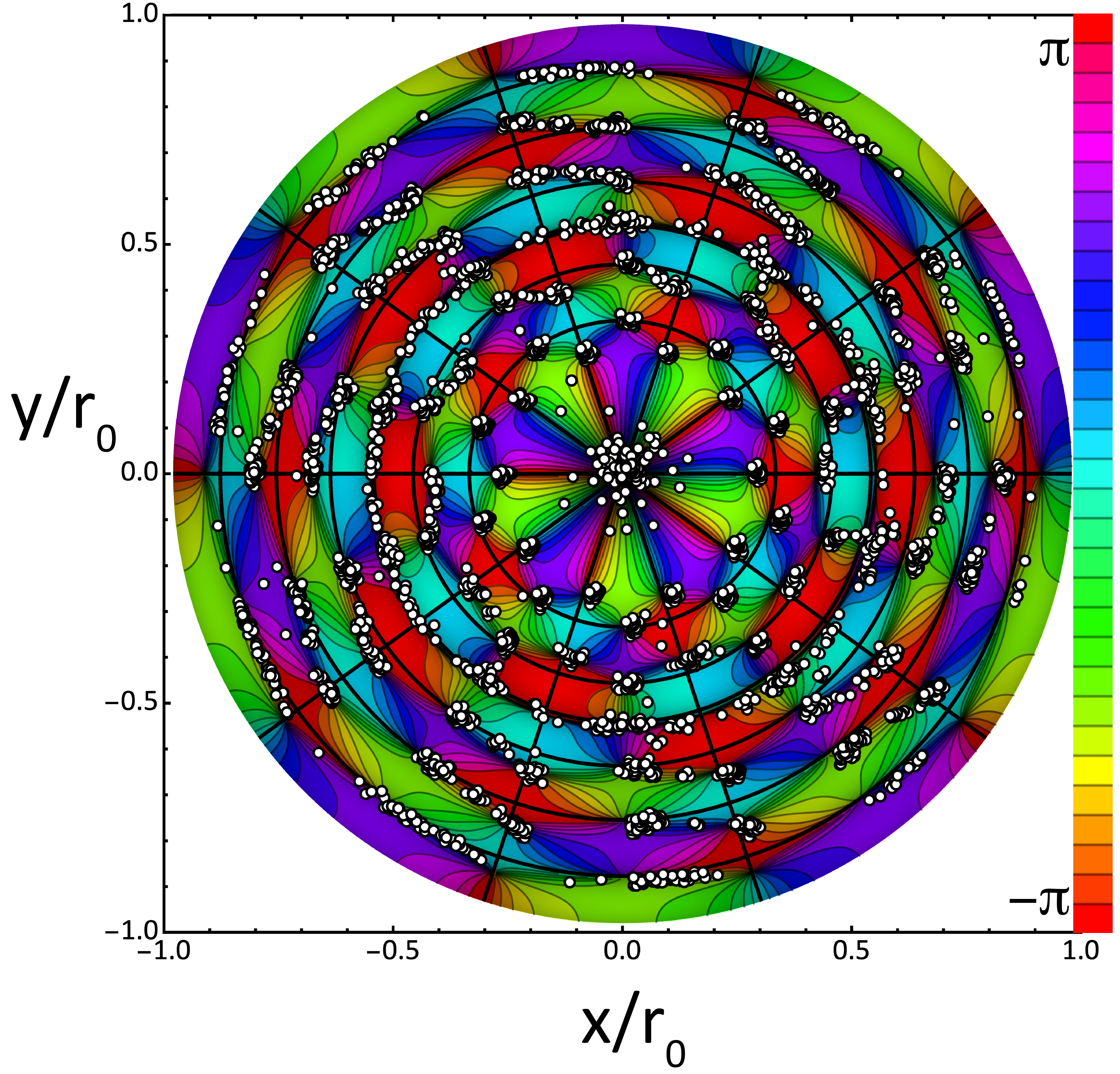}
	\end{center}
	\caption{ \emph{Large Vortex Arrays}.  An eigenstate of Eq. \ref{usol} is shown that is sufficiently large so that the polar-array structure is easily visible. Here $\beta=120^\circ$, $m=5$, $j_1 = 5$, $j_2 = 7$. White dots correspond to experimentally measured vortex positions, as discussed in the text, and correspond to a hard trap radius of $r_0 = 1.6$ mm.}
	\label{Large_Vortex_Array}
\end{figure}

Higher-order Bessel modes, $m$, elicit more complex polar arrays, as shown in Fig. \ref{Result_Composite_2}. Panel (a) amounts to a doubling of the single charges and dipoles observable in Fig. \ref{Dipoles_1}(a), while panel (b) is an example of how the mode parameters can be tuned to realize a specific array design. In this case, a mixed-charge pattern has been produced in which the charges fall on the vertices of equilateral triangles. A final array eigenstate is shown in Fig. \ref{Large_Vortex_Array} to give a sense for the general structure of higher-order vortex arrays.

%
%

%
%

	\section{Experimental Implementation}

\begin{figure*}[h]
\centering
\includegraphics[width=.8\linewidth]{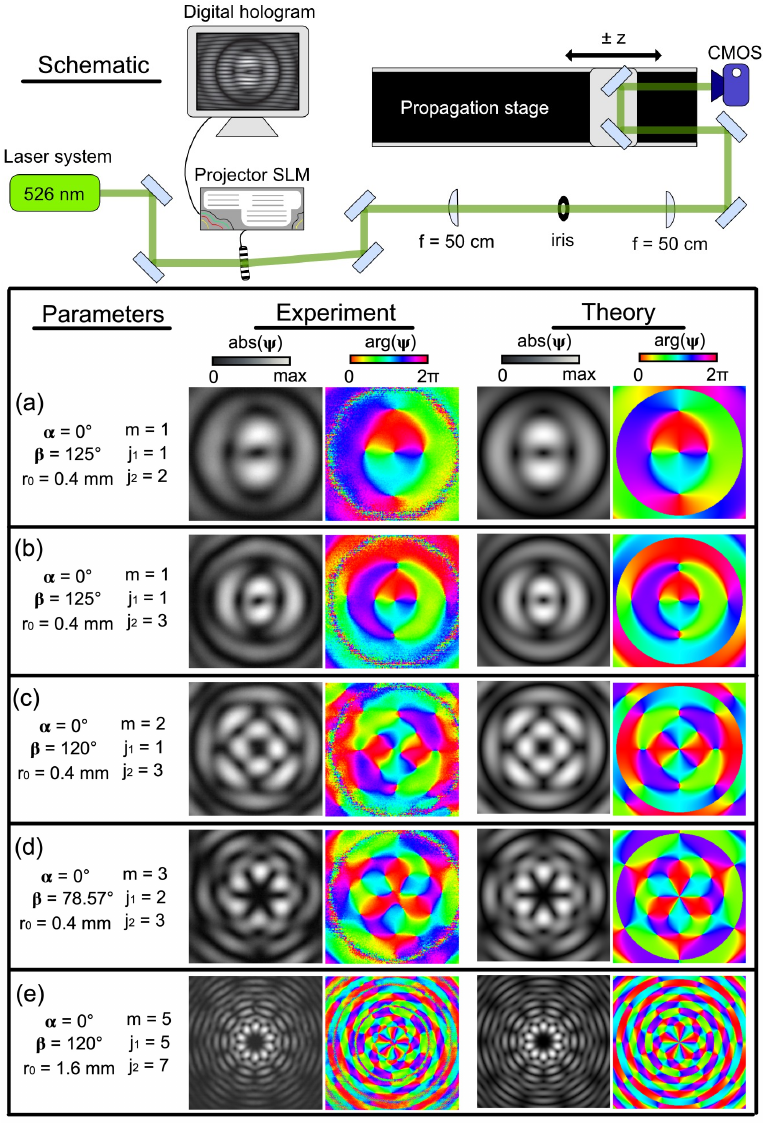}
\caption{(Top) Schematic of the experiment. Laser diode emission from a single-mode fiber is subsequently collimated. A computer controls the display of digital holograms on the SLM, sends commands to the motorized propagation stage, and controls the CMOS detector that captures intensity images. A 4f-imaging system, including two $50$ cm focal length lenses and an iris, selects the first diffracted order from the SLM and allows for data collection starting at the $z=0$ initial state. Only the first diffracted order from the hologram is depicted after the SLM. (Box) Comparisons between theoretical predictions and experimental implementations for all five examples. Transverse profiles, both amplitude ($\mathrm{abs}(\psi)$) and phase ($\mathrm{arg}(\psi)$), are generated/measured at $z=0$.}
\label{Fig5}
\end{figure*}

The examples of the previous sections were implemented and tested in a propagating laser beam, using the apparatus in the schematic of Fig. \ref{Fig5}. Vortex arrays were generated using a collimated, single-mode Gaussian laser beam ($\lambda = 526$ nm) that was transmitted through a spatial light modulator (SLM) \cite{Huang2012}. The SLM displayed a digital hologram, constructed by turning the amplitude and phase of the initial condition Bessel trap field, Eq. \ref{usol}, into a diffraction grating. This digital hologram is a two-dimensional discrete array matching the pixel pitch ($12.4$ microns) of the SLM and its resolution ($1024 \times 768$), and it has the following structural form:
\begin{align} \label{hologram}
    \mathrm{H}(x,y) = \frac{|u_{j_1,j_2}^{(m)}|}{\mathrm{max} \left(|u_{j_1,j_2}^{(m)}|\right)} & \times |0.5 \, e^{i \, \mathrm{arg} \left(u\right)} + \nonumber \\
    & 0.5 \, e^{i k_{\mathrm{g}} \left(\sqrt{3} x /2 + y / 2\right)} |.
\end{align}
The first term on the right hand side of Eq. \ref{hologram} controls the amplitude to match the target mode, and the next term specifies the phase. The last term puts the phase into a sinusoidal grating with
wavenumber 
$k_{\mathrm{g}} = 2\pi / N$, 
which defines the fringe separation 
($N=$5 pixels in these experiments) on the SLM. The scaling parameters in front of the $x$ and $y$ components of the planewave cause the grating to be ``diagonal'' so that the first-order diffracted beam avoids interference with pixel-diffracted beams. This hologram allows the construction of any vortex array, including complete experimental control over the Bessel trap parameters of $\alpha$, $\beta$, $m$, and $r_0$, by simply providing experimental values for them in Eq. \ref{hologram}. By propagating a Gaussian laser beam through the resulting hologram-modulated diffraction grating, as depicted in Figure \ref{Fig5}, the first-diffracted order is imprinted with the vortex array that was programmed into the hologram.

The first-diffracted order was isolated
using an iris at the focus of a 4f lens imaging setup, and then measured with a camera at the imaging location. Propagation of the mode was measured by stepping the propagation distance with a Newport M-IMS400PP translation staged. At each step, an intensity image of the beam was acquired with a WinCamD-LCM CMOS camera; this was converted to an amplitude image by taking a square root of each pixel. A phase image was acquired using collinear phase-shifting digital holography \cite{Andersen2019, Yamaguchi1997}; this entailed programming composite holograms, composed of a sum of the hologram of the mode to be measured with a zero-order Bessel mode of the same hard trap radius. The zero-order Bessel mode acted as the reference, and we made four new holograms with reference phase shifts of $0, \pi/2, \pi$, and $3\pi/2$. From camera measurements of the four phase-stepped interferograms, we then calculated the phase map of the signal field.

Fig. \ref{Fig5} compares predicted and measured amplitude and phase of Figs. \ref{Dipoles_1}-\ref{Large_Vortex_Array} for the initial $(z=0)$ transverse plane. Consistent with the experimental vortex positions shown in those figures, these magnitude and phase images shows excellent visual agreement in each case. The vortex positions featured as white dots in Figs. \ref{Dipoles_1}-\ref{Large_Vortex_Array} were measured by combining these amplitude and phase measurements into a single complex field for each $z$-step: $\psi = |\psi| \, e^{i \, \mathrm{arg}(\psi)}$. For each propagation step, we tracked the vortices of Figs. \ref{Dipoles_1} and \ref{Result_Composite_2} by finding the intersections of zero-crossings in the real and imaginary parts of the field $\psi$. Vortex tracking for Fig. \ref{Large_Vortex_Array} was accomplished by finding locations of minimum value in the amplitude. The modified approach of the latter case was called for because the array features a high density vortices and the associated phase gradients are severe \cite{Gorshkov2002}.

Some key differences are evident comparing theory to experiment, and these lend insight into the physics. First,  vortices at the trap center with a charge $|m| > 1$ are unstable and dissociate into $m$ unit-charge vortices. This well-known effect~\cite{Voitiv2022} is evident at the center of the experimental phase maps (colored with legend showing scale) of subpanels (c)-(e) of Fig. \ref{Fig5}. Second, the hard trap boundary is clearly visible in the experimental amplitude, but it is pixelated in the phase maps due to the low levels of light at those radial distances used in the phase calculation. Third, there are small deviations in the symmetry of the experimental profiles, as in the lack of symmetry with respect to the horizontal axis of the central vortex in Figs. \ref{Fig5} (a) and (b). Such small differences are typical and unavoidable for finite-aperture systems. 

There is a fourth key difference between the theory and the experimental demonstration: Bessel-Gaussian beams, rather than idealized Bessel modes, were generated \cite{Gori1987}. These beams have a finite lifetime---i.e. a maximum propagation distance---for which they can be reasonably approximated as non-diverging \cite{Durnin1987} and for which vortex dynamics will follow trajectories predicted in their idealized-Bessel counterparts \cite{Voitiv2020}. As a consequence, the Bessel hard trap does not persist indefinitely in the experiment. Vortex positions will therefore tend to drift as the beam diverges, an effect that is more pronounced with increasing distance from the center of the trap. This is clearly seen in Fig. \ref{Large_Vortex_Array}, by contrasting the discrete ``clumps'' of vortex positions in the center of the trap with the ``trails'' of vortex positions near the trap boundary. Nonetheless, within these modest experimental restraints, the implementation produced satisfying match with the proposal---demonstrating the major benefit of a simple, accessible experimental setting for these vortex array dynamics.

	\section{Conclusion}

The examples considered, along with others that were used to test our understanding, allow the following observations. For the sake of definiteness, we have restricted attention to cases for which $j_1 < j_2$. 
\begin{itemize}

\item There is a singularity of order m at the center of the trap.

\item  The parameter, $\alpha$, produces a rigid rotation of the array by $\alpha/2$ about the trap center.

\item The presence of vortices outside of the trap center requires that $j_1 \ne j_2$. Swapping $j_1$ and $j_2$ while also replacing $\beta$ with $\pi/2-\beta$ results in the same vortex structure.

\item Vortices lie at the intersection of rings and radial lines. There are $4m$ radial lines and a maximum of $2(j_2 - 1)$ rings. Each radial line contains a maximum $j_2 - 1$ vortices. 

\item There are only two types of radial vortex patterns, and they alternate azimuthally. For $\alpha = 0$, one type is along $\phi = 0$ while the other is along $\phi = \frac{3\pi}{2m}$. There are $2 m$ replicates of this pair of lines, rigidly rotated by an integer multiple of $\pi/m$.

\item Replacing $\beta$ with $2\pi - \beta$ swaps the vortex structure of the two types of radial lines.

\item For $\beta < \pi - (j_2 - 1) \frac{\pi}{j_2}$ or $\beta > \pi + (j_2 - 1) \frac{\pi}{j_2}$, there are at most $j_1 - 1$ vortices along each radial line. As $|\pi - \beta| \rightarrow 0$, the vortex cores become increasingly tilted and are extended azimuthally.

\end{itemize}

In conclusion, we have demonstrated that polar vortex arrays can be generated within an all-optical hard trap using a linear combination of just two Bessel modes. From the perspective of a steadily rotating frame, each of these is an eigenstate. The requisite rotation rates are quantized, calling to mind BEC vortices produced in a rotating trap\cite{Yarmchuk1979}, but the current setting is both linear and classical. Bessel traps should also prove useful in the study of confined vortex dynamics that are not stationary and can be used to produce annular domains with hard boundaries on either edge. 

The pseudo-potential associated with a rotating frame can be interpreted as a potential function that changes the free energy relative to eigenstates associated with a fixed frame, and the use of free energy in association with optical states is worthy of further investigation. 

The current work has focused on Bessel traps with a circular domain. However, the same approach can be immediately applied to consider annular domains as well. Such settings may exhibit their own novel features because there is no longer a trap center nor any vortex associated with this position.  It is also possible to construct analogous array eigenstates within a harmonic trap, but the setting is not as accessible since it requires that the light be propagating through an environment with inhomogeneous dielectric character.

\section{Acknowledgments}
The authors acknowledge support from both the W.M. Keck Foundation and the NSF (DMR 1553905).


\end{document}